\documentclass{aastex62}

\usepackage{threeparttable}
\usepackage{multirow}

\graphicspath{{./}{figures/}}

\shorttitle{The energetic thermonuclear bursts in SAX J1712.6--3739}
\shortauthors{Lin and Yu}

\begin{document}

\title{The energetic thermonuclear bursts in SAX J1712.6--3739}

\author[0000-0003-3965-6931]{Jie Lin}

\affiliation{Physics Department and Tsinghua Center for Astrophysics (THCA), Tsinghua University, Beijing, 100084, China: linjie2019@mail.tsinghua.edu.cn}
\affiliation{Shanghai Astronomical Observatory and Key Laboratory for Research Galaxies and Cosmology, Chinese Academy of Sciences, 80 Nandan Road, Shanghai 200030, China: wenfei@shao.ac.cn}
%\affiliation{University of Chinese Academy of Sciences, 19 A Yuquan Road, Beijing 100049, China}

\author{Wenfei Yu}
\affiliation{Shanghai Astronomical Observatory and Key Laboratory for Research Galaxies and Cosmology, Chinese Academy of Sciences, 80 Nandan Road, Shanghai 200030, China: wenfei@shao.ac.cn}

\begin{abstract}
The neutron star low-mass X-ray binary SAX J1712.6-3739 has been known for its long and hard thermonuclear X-ray bursts from previous observations. Its thermonuclear bursts are so distinct that they can last for tens of minutes as seen with \emph{Swift}/BAT ($E>15$~keV). To explore the origin of these extreme bursts and the nature of SAX J1712.6-3739, we analyzed the observations of all the four bursts which were captured by \emph{Swift}/BAT, and derived the peak flux and the fluence of these bursts from joint studies with \emph{Swift}/XRT and \emph{Swift}/BAT.
The derived bolometric peak fluxes observed by \emph{Swift} set the distance of SAX J1712.6--3739 to be 4.6--5.6~kpc, while the derived absolute magnitude and average accretion rate agrees with its ultra-compact nature. Our measurements of the effective duration of these bursts conclude that the 2010 burst corresponds to a normal X-ray burst, the 2011 burst is consistent with an intermediate-duration burst, while the 2014 and the 2018 bursts are more energetic than common intermediate-duration bursts but less energetic than those known superbursts. We estimated that the average mass accretion rate of SAX J1712.6--3739 was about only 0.4--0.7\%~$\dot{\rm M}_{\rm Edd}$. Current theory predicts no carbon production in the bursters under such low accretion rate. If true, the 2014 and the 2018 bursts are then deep helium bursts instead of carbon bursts. The thermonuclear bursts of SAX~J1712.6--3739 have shown a very wide range of duration. The ignition model predicts that the diverse burst durations are induced by variable accretion rates, but current results provide only weak support to this inference.
\end{abstract}

\section{Introduction} \label{sec:intro}

X-ray burst phenomenon in the neutron star (NS) X-ray binaries (XRBs) is produced by the thermonuclear burning runaway on the surface material of NS. 
Hence, such thermonuclear bursts are used as one of the key diagnostics to distinguish neutron stars from black holes in XRBs. Based on the duration and the integrated radiative energy measured in X-ray observations, thermonuclear bursts are divided into three categories, namely normal X-ray bursts with duration of tens of seconds, intermediate-duration bursts with a duration of tens of minutes, and superbursts with a duration of a few hours. Classifications based on other burst properties also exist. During some of the thermonuclear bursts, strong radiation pressure leads to the photospheric radius expansion (PRE) with a sudden drop of effective temperature. In the PRE stage, the burst radiative luminosity reaches the local Eddington limit. It is expected to remain at almost constant level until reaches the ``touch-down'' point when the neutron star atmosphere reaches the surface again \citep{Kuulkers+etal+2003}. These PRE bursts therefore have been used as ``standard candle'' to determine the distances of these X-ray bursters. Interestingly, in a small fraction of PRE bursts, some extreme PRE processes makes the temperature so low that the X-ray emission shifts to the UV wavelengths for a few seconds. Hence, in the X-ray light curves of these so-called ``superexpansion'' bursts, a ``precursor'' prior to a main burst is usually shown \citep{Kuulkers+etal+2003,Zand+weinberg+2010}.

In the theory of burst ignition, the thermonuclear runaway occurs when the radiative cooling cannot moderate the small perturbations in the heating generated from the nuclear reactions.
And the ignition conditions are very sensitive to the accretion rate, since the energy generation due to compression is proportional to the accretion rate (see \citealt{Cumming+etal+2006,Galloway+Keek+2017}).
In general, for a lower accretion rate, the thermonuclear burning runaway is more likely to be triggered at a higher column depth. Therefore, the burst energies of a neutron star could be various if its accretion rate is significantly varied with time.

SAX J1712.6--3739 is a persistent low-mass X-ray binary discovered by BeppoSAX/WFC \citep{Zand+etal+1999+discovery,Cocchi+etal+2001}, which was also detected by ROSAT survey as 1RXS J171237.1--373834 (see \citealt{Wiersema+etal+2009}). From the BeppoSAX/WFC observation in Spetember 1999, \cite{Cocchi+etal+2001} suggested that the distance to SAX J1712.6--3739 is about 7 kpc by assuming that the bolometric peak flux $F_{\rm peak} \approx 5\times10^{-8} \rm \,ergs\,cm^{-2}\,s^{-1}$ is very close to the Eddington luminosity. 
Furthermore, due to its very low average accretion rate, \cite{Zand+etal+2007} suggested that SAX J1712.6--3739 is an ultra-compact X-ray binary (UCXB) candidate, and they predicted that the visual magnitude $m_{\rm V}\approx26$, which may be used to confirm its UCXB nature by assuming the visual extinction $A_{\rm V}=7.3$ and the distance $d=7$ kpc. However, from the observation performed by the EFOSC2 instrument on the ESO 3.6-m telescope, the visual magnitude of SAX J1712.6--3739 was $m_{\rm V}=24$ \citep{Wiersema+etal+2009},  which leads to an absolute magnitude of $M_{\rm V}\approx2.4$, somewhat brighter than the absolute magnitude $M_{\rm V}\approx3.7-5.6$ for typical UCXBs (see \citealt{Paradijs+McClintock+1994}).  

A dozen thermonuclear bursts including quite a few long and energetic bursts in SAX J1712.6--3739 have been observed by BeppoSAX/WFC\citep{Cocchi+etal+2001,MINBAR+2020}, RXTE/ASM \citep{Kuulkers+2009+atel}, 
INTEGRAL \citep{Chelovekov+etal+2006,Alizai+etal+2020} and \emph{Swift} \citep{Strohmayer+Baumgartner+2010+atel,Palm+2011+atel,Cummings+etal+2014+GCN,Lin+Yu+2018+atel,Iwakiri+etal+2018+1712}. \cite{zand+etal+2019} found that the 0.8-hour X-ray burst observed in August 2014 (hereafter B2014) was the longest burst in their samples from \emph{Swift} observations. Since the burst duration is comparable to that of the shortest superburst now known and also that of the longest intermediate-duration burst, whether B2014 is a superburst or an intermediate burst is not determined. In this paper, we intend to investigate the observational properties of the four bursts captured by the wide field of view (FoV) of \emph{Swift}/BAT, incuding B2014, to exploring the natures of SAX J1712.6--3739 and its amazingly long and energetic thermonuclear bursts.

\section{Data Reduction} \label{sec:data}

\subsection{\emph{Swift}/BAT observations of bursts}

The \emph{Swift}/BAT can record several data products by the flight software on board. In response to a trigger by a gamma-ray burst or a special X-ray burst event, the software will produce and store BAT \emph{event data}, contain complete time and energy channel information of each incident photon. The BAT event data allows us to extract final data products (such as light curves and spectra) in certain temporal re-binning for any source position in the FoV.
However, due to deficiency of the on-board storage and downlink capacity, BAT event data are only available for special triggers. Usually these observation data is collected into detector plane histograms (DPHs) with a typical time resolution of 5 minutes. The DPHs data (namely \emph{survey data}) covers much longer period of time than the BAT event data does. it contains an 80-channel spectrum for each detector and can be used to extract data products for sources in any positions in the FoV, although the survey data cannot be used to study variations or phenomena shorter than 5 minutes. Furthermore, the flight software also produces rate data all the time, which is generally used for the searching of triggers and the monitoring of the instrumental state. The \emph{rate data} contains only summed counts of detector units and thus cannot provide final data products for specific source in the sky. 

\begin{figure}
\plotone{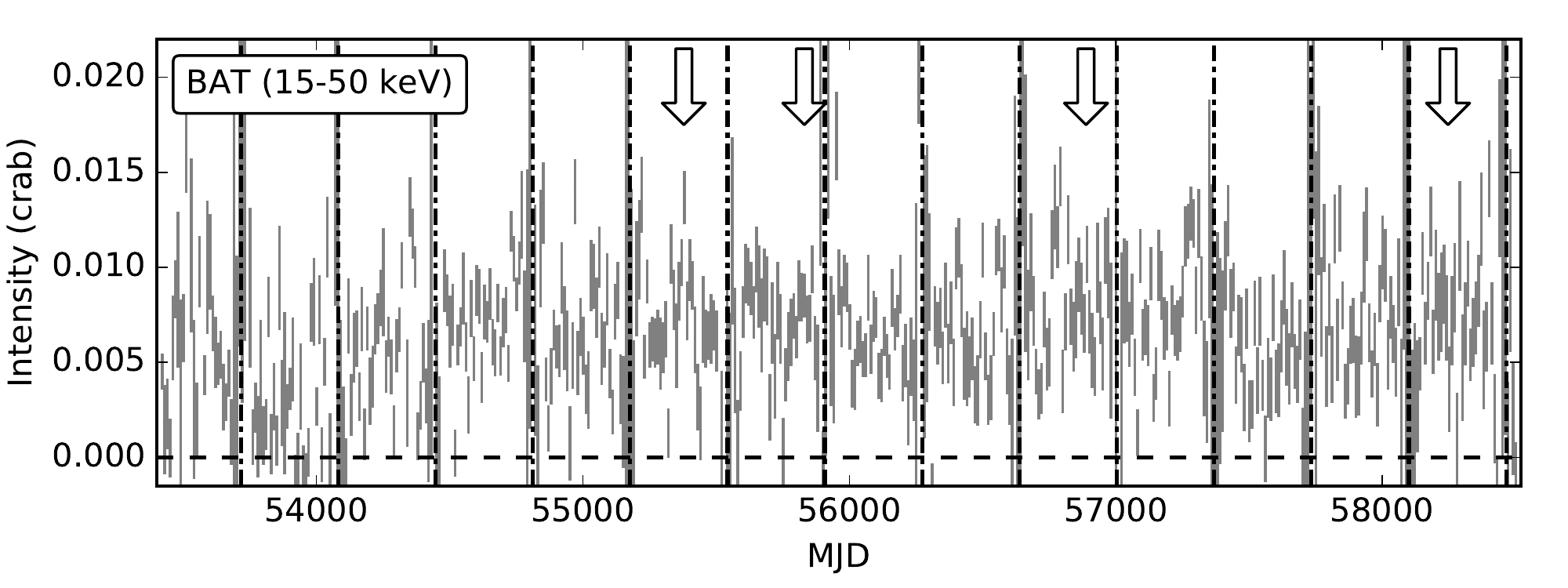}
\caption{The \emph{Swift}/BAT light curve of SAX J1712.6--3739 from 2005 February to 2019 January. The count rates were rebinned into 10-day average rates and then converted into crab unit by $\rm 1\, crab = 0.22\, counts\,cm^{-2}\,s^{-1}$.
The arrows indicate the time corresponding to the X-ray bursts captured by BAT observations, and the dotted-dashed lines point the time when angle separation between SAX J1712.6--3739 and sun reaches the minimum value, 14.6 degrees.
\label{figure_persistent}}
\end{figure}

As Figure~\ref{figure_persistent} shows, four thermonuclear bursts of SAX J1712.6--3739 have been captured by \emph{Swift}/BAT. These bursts were detected in September 2010 (hereafter B2010, \citealt{Strohmayer+Baumgartner+2010+atel}), September 2011 (hereafter B2011, \citealt{Palm+2011+atel}), August 2014 (B2014, \citealt{Cummings+etal+2014+GCN}) and May 2018 (hereafter B2018, \citealt{Lin+Yu+2018+atel,Iwakiri+etal+2018+1712}), respectively.
Among them, B2010, B2011 and B2014 triggered recording in BAT event mode, which were followed by automatic pointed X-ray observations with the XRT (see also \citealt{zand+etal+2019}). However, the most recent burst B2018 did not trigger, so there were no BAT event mode data nor follow-up XRT pointed observations, thus we investigate B2018 using its BAT survey data. In particular for the 2014 burst event, because there is a $\approx 400$~s gap in the BAT event data, we investigated its BAT survey data and rate data during the gap. 

We analyzed the BAT event data following the standard analysis threads introduced by UK \emph{Swift} Science Data Centre \footnote{\url{https://www.swift.ac.uk/analysis/bat/index.php}} .
Since the BAT team did not support general survey data analysis, we analyzed the survey data following a similar procedure to the event data. 

To ensure the correct energy conversion from the ADU values to the energies in the event data files, the BAT event data was first corrected by \emph{HEAsoft} BAT Energy Conversion tool \emph{bateconvert}, which can fill the PI column in the event files.
On the other hand, the survey data already has a crude energy correction applied on-board. But for getting more exact energies, we ran BAT Energy Rebin task \emph{baterebin} and applied the gain/offset maps generated by BAT flight software to correct the survey data.

We extracted the Detector Plane Images (DPIs) from the event (or survey) data by \emph{batbinevt} task,
and then ran the task \emph{bathotpix} to locate good pixels in the DPIs and produce detector quality map.
The DPI file is a map of the counts (or count rates) for each detector, which is the ``starting point'' to reconstruct the sky fluxes from detector data. The detector quality map is used for computing the fraction of illuminated pixels, which contributes to the computation of mask weighting pattern.
We computed the mask weighting map corresponding to the position of SAX J1712.6--3739 and overwrote the MASK\_WEIGHT column in the BAT event files by running Mask Weights Computation task \emph{batmaskwtevt}.
For BAT survey data, we constructed mask weighting map files by applying another Mask Weights Computation tool \emph{batmaskwtimg}.

We used \emph{batbinevt} task to extract the final data products (light curves and energy spectra) for both the event mode and the survey mode data.
The \emph{batbinevt} task can use the mask weighting pattern in a MASK\_WEIGHT column or a mask weighting map file to output the light curves and spectra specific to our time range selection.
Furthermore, the systematic error vectors, obtained from BAT calibration database in CALDB, were added into the energy spectra by the tool \emph{batphasyserr}.
The BAT ray tracing columns in the spectra were updated by tool \emph{batupdatephakw} to avoid  erroneous detector response matrix due to the slew of spacecraft. Finally we produced the detector response matrix files (RMFs) by the BAT Detector Response Matrix (DRM) generator tool \emph{batdrmgen}.

The Mission Elapsed Times (METs) recorded in the raw FITS files are in the time standard of Terrestrial Time (TT). We corrected the METs by the the Universal Time correction factor (UTCF) and leap seconds, and then converted them into the dates in the time system of Coordinated Universal Time (UTC).
For these BAT light curve products, we set the zero points to the time when the burst flux significantly increased over the persistent flux. Hence, the start times were slightly different from the BAT trigger times \citep{Strohmayer+Baumgartner+2010+atel,Palm+2011+atel,Cummings+etal+2014+GCN}.

It is worth noting that, in the \emph{Swift}/BAT light curves (see the panel b of Figure~\ref{figure_lightcurve}), there was a ``precursor'' prior to a main burst during B2014.
But unfortunately, the BAT event data and the BAT survey data did not cover the rise stage of the main burst of B2014.
Because \emph{Swift} did not slew between the trigger of BAT event-data recording and the end of XRT pointed observation, we can investigate the BAT rate data to reproduce the rise stage of the main burst. 
We extracted the 15--100 keV counts from the 64-millisecond rate file of the \emph{Swift} observation (ID: 00609878000). The raw counts were rebinned to the unite of counts per 2.048s, which was consistent with the light curves extracted from the event data, and then converted to count rates. 
The net rates was obtained by the subtraction of the 300s-average pre-burst flux.  
To reduce the count rate for the projection effect, we divided the net count rate by the cosine of off-axis angle.
Then to further calibrate the net count rates, we produced a mask weighting image corresponding to the position of SAX J1712.6--3739. We read the number of enable detectors and the partial coding fraction from the header of the image file, and then estimated an average illumination fraction for the illuminated detectors from the table of the image.
The net rates were finally normalized to the total illuminated areas of enable detectors.

\subsection{\emph{Swift}/XRT observations}

Due to high X-ray count rates, these X-ray bursts of SAX J1712.6-3739 were mainly observed automatically in the window timing (WT) mode with \emph{Swift}/XRT. We reduced the XRT data following the standard data analysis threads for the WT mode \footnote{\url{https://www.swift.ac.uk/analysis/xrt/index.php}}. The data was all calibrated by running the \emph{HEAsoft} tool \emph{xrtpipeline}.
We used the task \emph{Xselect} to extract the light curves from the clean event files. 
To avoid pile-up effect, the extraction region for the source was set to an annullus with an inner (outer) radius of 2 (40) pixels, and the region for the background was set to an annulus with an inner (outer) radius of 60 (120) pixels. These light curves were all corrected for telescope vignetting and point spread function (PSF) by the task \emph{xrtlccorr}.

Furthermore, in order to estimate the peak flux and the fluence of these bursts, we produced the time-resolved energy spectra from the XRT data. We divided the clean event files into multiple  sub-event files with 700 XRT counts in each time slice, and then extracted the energy spectra from these sub-event files by \emph{Xselect}. Due to the large span of the count rates during these bursts and severe pile-up effect in those sub-event files of high count rates,
the source extraction region was set to a circle with a radius of 40 pixels if the count rate was below 100 counts s$^{-1}$, or an annulus with an inner (outer) radius of 2 (40) pixels if the count rate was above 100 counts s$^{-1}$. The background spectra were extracted from an annulus with an inner radius of 60 pixels and an outer radius of 120 pixels. 
Noticed that, since SAX J1712.6--3739 is heavily absorbed ($N_{\rm H} \gtrsim 10^{22}~{\rm cm^{-2}}$), we selected only grade-0 photons for these spectra ( because of the low energy spectral residuals in event grades 1 and above, see also \url{https://www.swift.ac.uk/analysis/xrt/xselect.php }) and then grouped them into 20 counts per spectral bins. We applied the corresponding response matrix files (RMFs) from the Swift/XRT calibration database in CALDB and produced ancillary response files (ARFs) from exposure maps using the \emph{xrtmkarf} task.

The spectra of pre-burst or post-burst emission was not subtracted from the resulting spectra, because the persistent fluxes were two orders of magnitude lower than the burst flux at least (see also \citealt{zand+etal+2019}), and the persistent flux could vary during bursts due to dramatically increase of seed photons. Subtraction of the pre-burst or post-burst flux may not completely remove the contribution from the accretion disk or corona (see \citealt{Degenaar+etal+2018}).

We then fit these time-resolved energy spectra in 0.5--10 keV band with an absorbed blackbody model, where the interstellar absorption was accounted by WABS model \citep{Morrison+McCammon+1983}. 
From the best-fit blackbody model, we derived the unabsorbed bolometric flux at 0.1--200 keV for all spectra. The quoted errors represent 1$\sigma$ confidence intervals throughout this paper.

\section{Results} \label{sec:result}

\subsection{Light curves}
\begin{figure}
\plotone{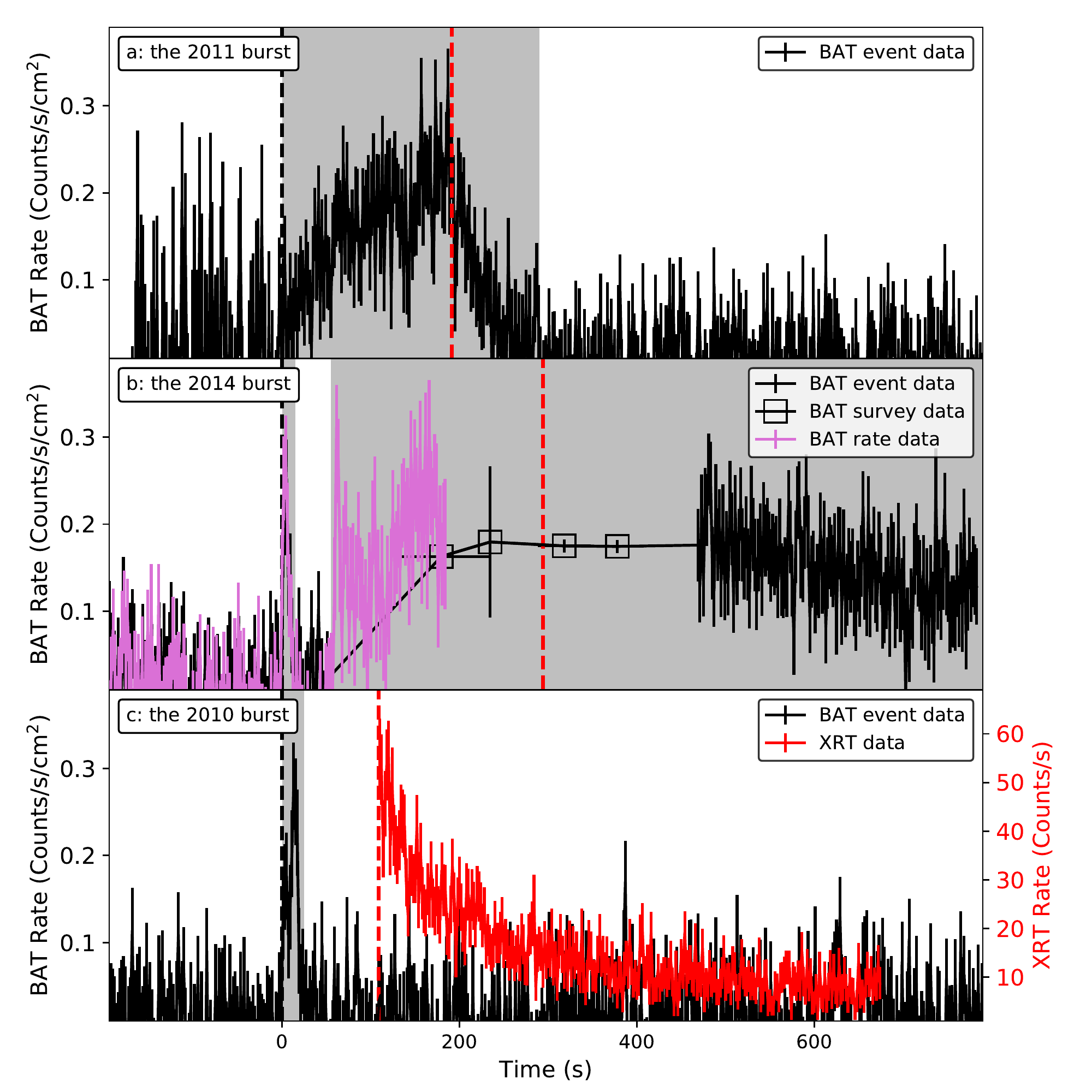}
\caption{The light curves of the 2011 (panel a), the 2014 (panel b) and the 2010 (panel c) burst events with a typical time binning of 2 seconds. 
All three X-ray bursts have triggered automatic XRT observations. In order to illustrate that the 2010 burst event was actually much more powerful than it appeared in the BAT light curve, its XRT light curve is plotted together.
The black and red vertical dashed lines indicate the start times of the bursts and the start times of the XRT observations, respectively. The grey areas are used to mark the periods when the bursts were detectable with the \emph{Swift}/BAT observations. In particular, for the 2014 burst event, the light curves were generated from the BAT event data (black solid line), BAT survey data (black square) and BAT rate data (purple solid line).
\label{figure_lightcurve}}
\end{figure}

\label{sec:light_curve}

We show the X-ray light curves of B2011, B2014 and B2010 in a typical time bin of 2 seconds in the Figure~\ref{figure_lightcurve}. B2011 (in panel a) started on UTC 2011 September 26 20:10:19 (as $T_0$), slightly earlier than the BAT trigger time 20:11:28 \citep{Palm+2011+atel}.
It was rising slowly as compared to other X-ray bursts in SAX J1712.6--3739, and reached its peak flux at $ \approx 0.3 \rm \,counts\,s^{-1}\,cm^{-2}$ ($\approx 1.4 \rm \,Crab$) at $T_0+175$s, and then fell below the detection limit at $T_0+290$s. Notice that, the start time and the duration mentioned here depend on instrumental sensitivity, the energy band and the time bins we chose.

As introduced in Section~\ref{sec:data}, the light curves of B2014 were produced from the BAT event, survey and rate mode data together. Its light curve (in panel b) generated from the rate data agrees with that of the event data, implying that the specific rate data is reliable for any further analysis. In the hard X-ray band 15--100~keV, we found there was a $\approx 15$~s ``precursor'' started on UTC 2014 August 18 17:12:12. After a $\approx 40$~s quiescence, the source count rate rose again and reached the peak value of $\approx 0.3 \rm \,counts\,s^{-1}\,cm^{-2}$ within only 3 seconds. The main burst lasted for more than 720 seconds at an almost constant flux plateau. Unfortunately Swift/BAT slew away so did not cover the source during the decay of B2014.

\begin{figure}[!htbp]
    \includegraphics[width=0.49\textwidth]{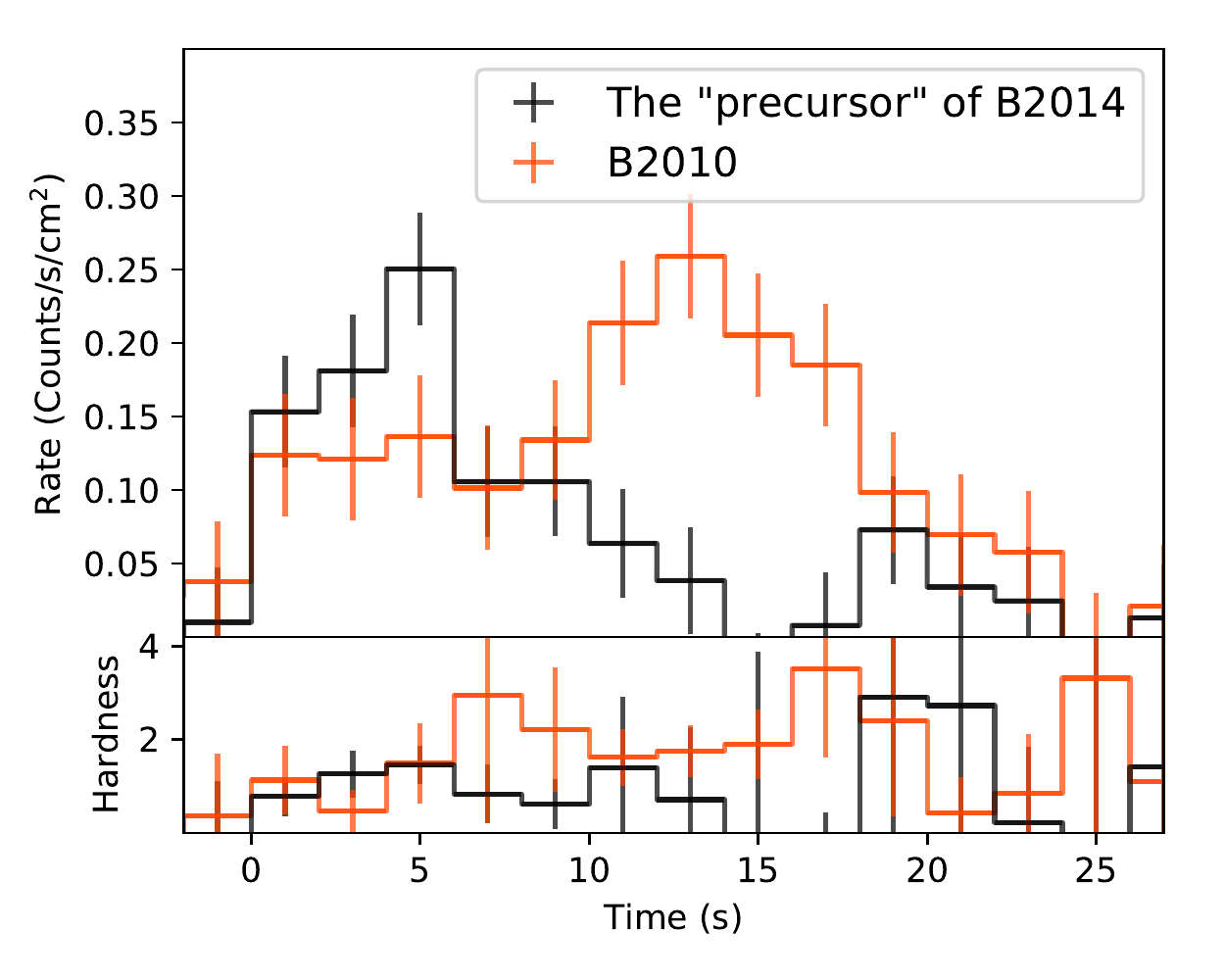}
    \includegraphics[width=0.49\textwidth]{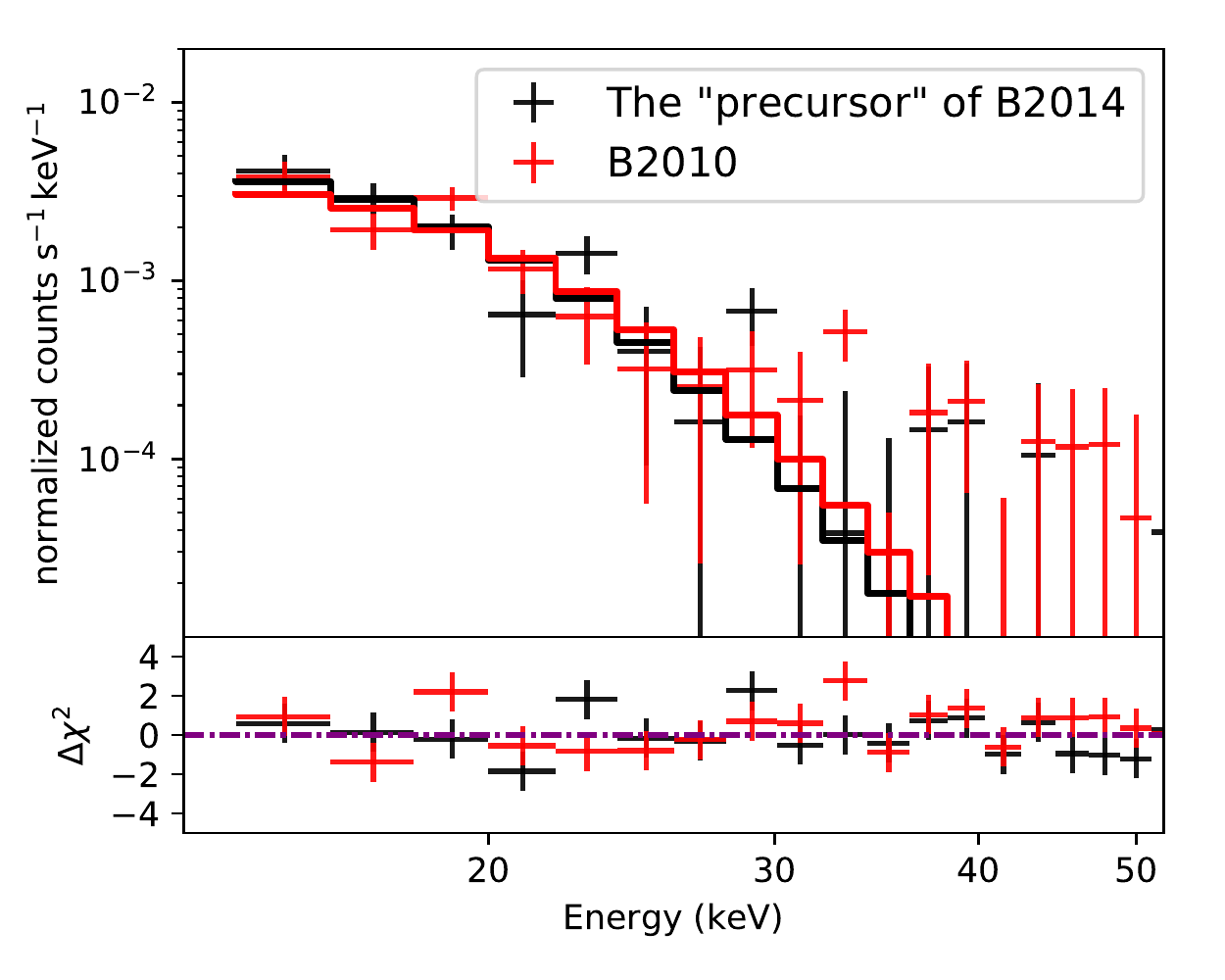}
    \caption{
     The comparison between the 2010 burst and the ``precursor'' of the 2014 burst.
     The left panel shows the zoom-in BAT light curves in the energy band of 15--50 keV and the hardness ratios between 18--50 keV and 15--18 keV. Their time bins correspond to 2 seconds.
     The right panel shows the BAT energy spectra and their best-fitting models, in the energy band from 15 to 50 keV.} 
    \label{fig:precursor_compare}
\end{figure}

For a detailed comparison, we show the BAT light curves of B2010 in the panel c of Figure~\ref{figure_lightcurve}. Its BAT light curve displayed a twin-peak profile and a duration of 20--25s, which was slightly longer than the ``precursor'' of B2014 but still comparable (see also Figure~\ref{fig:precursor_compare}).
It is worth noting that, the burst light curves as seen in BAT observations can be very different from those seen in 1--10 keV band with the XRT. A non-detection in the BAT observation does not mean that an X-ray burst is weak. On the other hand, it can be bright but just caused by a slight drop in effective temperature of the blackbody radiation. To show and illustrate that the 2010 burst was actually much more energetic than it appeared in the BAT light curve, the XRT light curve is also plotted at the panel c of the figure. 

Since the Figure~\ref{figure_lightcurve} shows that the ``precursor'' of B2014 had a similar duration and peak count rate to those of B2010, we tried to compare their hardness ratios and energy spectra in their BAT observations. 
In order to select roughly equivalent number of photon counts from the hard and the soft bands, we took 18--50 keV and 15--18 keV bands to calculate the hardness ratios. As shown in the left panel of Figure~\ref{fig:precursor_compare}, B2010 continuously harden until $T_0+17$s while the ``precursor'' of B2014 showed significant hardness ratios just before $T_0+10$s. The poor data quality does not allow us to further investigate the evolution of their spectral energy distributions, and thus we can not confirm whether the radiative radius $R$ is blowing up or not at the end of the ``precursor''.
Instead, we showed their integral energy spectra in the right panel of Figure~\ref{fig:precursor_compare}.

Their energy spectra were fitted by an absorbed black body model (wabs*bbodyrad) with a fixed absorbed hydrogen column density of $\rm 1.54\times10^{22}\,cm^{-2}$ \citep{zand+etal+2019}. As the results, the best-fitting effective temperatures and normalization were $2.61^{+0.35}_{-0.31}$ keV and $78.4^{+118.1}_{-45.1}$ for the ``precursor'', 
$2.89^{+0.35}_{-0.30}$ keV and $36.5^{+41.7}_{-19.1}$ for B2010, respectively.
It can be seen that the difference between their blackbody temperatures or radiative radii is not significant. 
And their 15--50 keV fluxes are consistent, as $6.3\times10^{-9}\rm\,ergs\,s^{-1}\,cm^{-2}$. Hence, their energy spectra are very similar.

\subsection{The estimates of the burst fluence and the effective duration}

The effective duration $\tau=f_{\rm b}/F_{\rm peak}$ is usually used to distinguish the three burst branches, where $F_{\rm peak}$ is the bolometric peak flux and $f_{\rm b}$ is the bolometric fluence of the burst, which is usually estimated by integrating all flux measurements for the burst. However, a large number of intermediate-duration bursts and superbursts are not completely covered by X-ray observations. In order to estimate their fluences, we could apply a broken power-law model to fit the burst flux as a function of time, and then obtain the peak flux and the fluence from the best-fitting model.

\begin{figure}[!ht]
\plotone{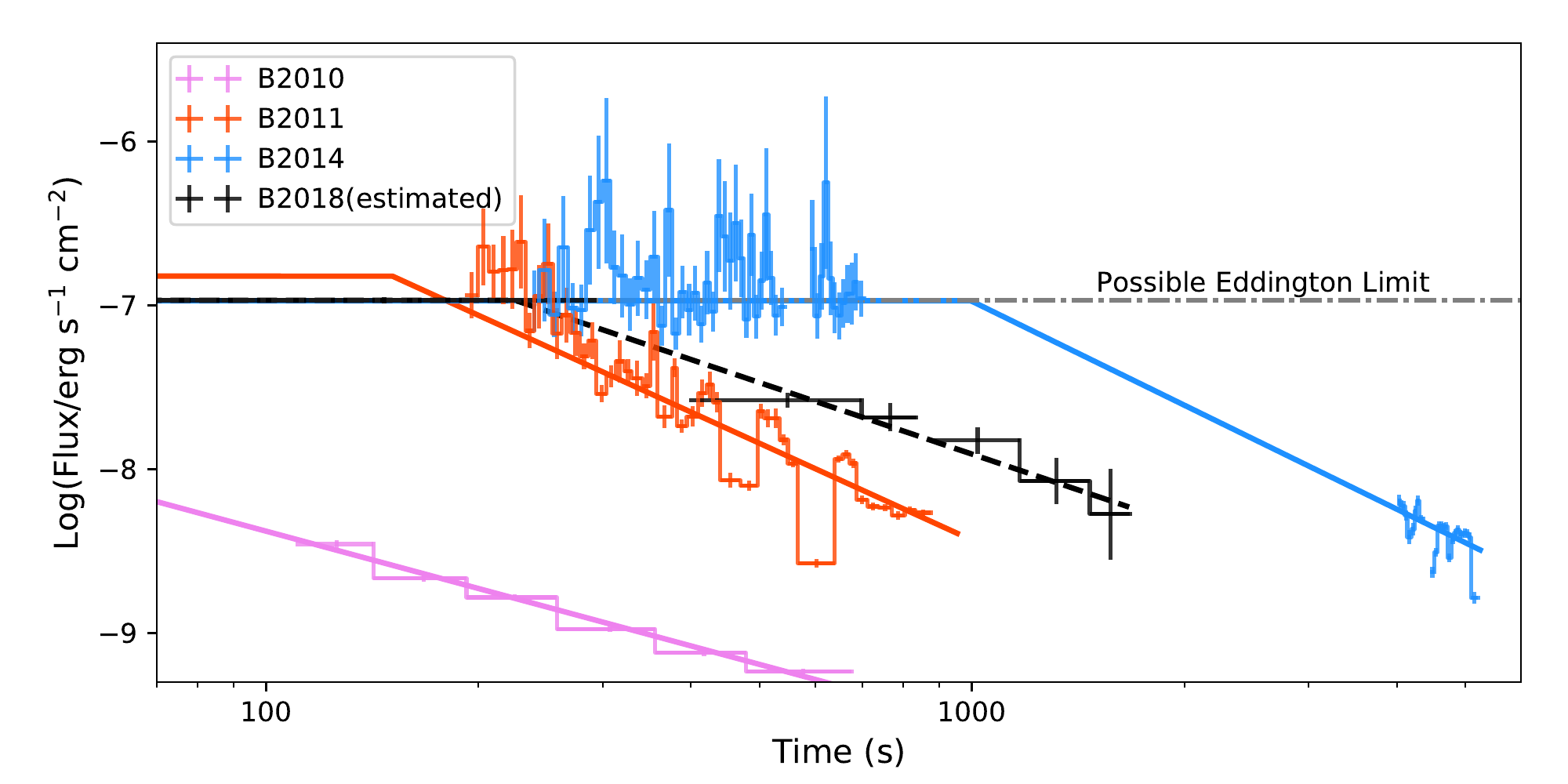}
\caption{The bolometric fluxes (0.1--200 keV) for the 2010, 2011 and 2014 burst events. The bolometric fluxes are extrapolated from the time-resolved XRT spectra (0.5--10 keV).
The start point of B2010 and B2011 here were set to the time when the burst flux was significantly detected from \emph{Swift}/BAT, while the start time of B2014 here was set to UTC 17:13:10 when the main burst just started.
Since the XRT observation did not cover the 2018 burst event and its start point cannot be determined, we roughly estimate the fluxes of the 2018 burst event by its BAT count rates (15--100 keV) from the survey data (DPHs files) and set the start time to UTC 2018 May 08 09:17:48 when SAX J1712.6--3739 entered the FoV of \emph{Swift}/BAT.
Their profiles are fitted with specific broken power-law models,  respectively. The dashed line indicates that the 2018 burst event is essentially stronger than the best-fitting model shows. 
The dot-dashed line represents the flux corresponding to possible Eddington limit. 
\label{figure_flux}}
\end{figure}

Among these burst events, B2010, B2011 and B2014 have been partially covered by \emph{Swift}/XRT, thus their burst fluences can be estimated from pointed observation data. 
As Figure~\ref{figure_flux} shows, B2014 kept a constant flux for about 550 seconds until the slew of \emph{Swift} interrupted the observation. After about 1 hour, when \emph{Swift} pointed to SAX J1712.6--3739 again, its flux decayed to below $10^{-8}\rm\,ergs\,s^{-1}\,cm^{-2}$.
Notice that, the start point of B2010 and B2011 here were set to the time when the burst flux was detectable from \emph{Swift}/BAT, while the start time of B2014 here was set to UTC 17:13:10 when the main burst just started.
B2014 and B2011 have an almost consistent peak flux, but XRT observation fail to cover similar ``constant stage'' for B2011. 

\begin{table*}[]
\caption{The best-fitting parameters and derived properties for four thermonuclear bursts in SAX J1712.6--3739.}
\label{table_result}
\renewcommand\arraystretch{1.2}
\footnotesize
 \begin{threeparttable}
\begin{tabular}{cccccccc}
\hline
 & $t_{\rm touchdown}$ &$\alpha$ & Peak Flux & Fluence & Effective Duration & Radiative Energy &Ignition Depth \\
&s&& $10^{-8} \rm\, ergs\,cm^{-2}\,s^{-1}$&$10^{-6} \rm\, ergs\,cm^{-2}$&  s &$10^{40} \rm\, ergs$ & $10^{8} \rm\, g\,cm^{-2} $\\
\hline
%\multirow{2}{*}{B2010}   & 111 $\pm$ 12 & 1.16 $\pm$ 0.04 &  0.37 $\pm$ 0.05 & 1.7 $\pm$ 0.1   & 471 $\pm$ 50 &  0.43 &  3.6 \\
\textbf{B2010}  & 6.1 $\pm$ 0.7 & 1.16 $\pm$ 0.04 &  10.7 (fixed)  & 3.5 $\pm$ 0.1   & 32 $\pm$ 1 &  0.88 &  7.3 \\
\textbf{B2011} & 151 $\pm$ 32 & 1.96 $\pm$ 0.05 &15.1 $\pm$ 6.5 &  45 $\pm$ 10 & 305 $\pm$ 65 & 11 & 94 \\
\textbf{B2014} & 996 $\pm$ 55 & 2.11 $\pm$ 0.08 &10.7 $\pm$ 0.6  & 195 $\pm$ 10 & 1826 $\pm$ 59 & 49 & 407 \\
\textbf{B2018} & $\gtrsim$225& $\gtrsim$1.45 & $\approx$10.7 & $\gtrsim$68.2& $\gtrsim$ 637 & $\gtrsim$17 & $\gtrsim$143 \\

\hline
\end{tabular}

\begin{tablenotes}
\item{\textbf{Notes.} The error here represents an uncertainty of 1$\sigma$. 
The fluxes of the 2010 burst event were modeled with a fixed peak flux.
Since the start time of the 2018 burst event cannot be determined, we showed the lower limit for its values here. 
The burst energy and ignition depths are calculated by assuming that the source distance is 4.6~kpc$\times\xi_{\rm b}^{-\frac{1}{2}}$ (see Section~\ref{sec:basic_natures}).
Additionally, the ignition depths here are calculated by assuming pure helium fuel (i.e. $Q=1.31\,{\rm MeV\,nucleon^{-1}}$, see Section~\ref{sec:disscussion:fuel}) and $R_{\rm NS}=10$~km. 
}
\end{tablenotes}
\end{threeparttable}
\end{table*}

The flat profile observed in the early phase of burst is usually thought to correspond to the PRE phase where the Eddington luminosity is reached \citep{Lewin+etal+1993,Kuulkers+etal+2003,Keek+etal+2017}.
We assumed that B2014 and B2011 both reached the Eddington luminosity and their burst profiles can be fitted with a broken power law (see also \citealt{Keek+etal+2017}),
\begin{eqnarray}
F(t)  =  \left\{
\begin{array}{lcr}
F_{\rm peak} && t\leq t_{\rm touchdown}\\
F_{\rm peak} \left (\frac{t}{t_{\rm touchdown}} \right)^{-\alpha} && t > t_{\rm touchdown} 
\end{array}\right. ,
\end{eqnarray}
where $F_{\rm peak} $ is the bolometric peak flux, and $t_{\rm touchdown}$ represents the ``touch-down'' time ( \citealt{Kuulkers+etal+2003,zand+etal+2019}). 
The burst fluxes and their best-fitting models are shown in Figure~\ref{figure_flux}.
It is worth noting that, there are some ``slow dips'' and ``fast fluctuations'' (see \citealt{zand+etal+2019}) in B2011 and B2014, which can also be seen in the long thermonuclear bursts of other bursters, e.g. 2S 0918-549 and IGR J17062-6143 \citep{zand+etal+2005,zand+etal+2011,Degenaar+etal+2013,Degenaar+etal+2018}.
%Because the early phases of B2010 and B2011 are not well covered by XRT, their peak fluxes cannot be constrained tightly. 

By calculating the weighted mean of peak fluxes of B2011 and B2014, we estimated the bolometric flux corresponding to Eddington luminosity (hereafter Eddington flux) as $F_{\rm Edd}= 1.07\pm0.05 \times 10^{-7}\rm \,ergs\,s^{-1}\,cm^{-2}$. %notice
As the panel c of Figure~\ref{figure_lightcurve} shows, the XRT observation did not cover the peak of B2010 at all, thus the best-fitting model gave an essentially underestimated peak flux $F_{\rm peak}=3.7\pm0.5 \times 10^{-9}\rm \,ergs\,s^{-1}\,cm^{-2}$ and an overestimated ``touch-down'' time $t_{\rm touchdown}=111\pm12$~s.
It is worth noting that, B2010 showed a peak BAT count rate similar to B2011 and B2014 (see Figure~\ref{figure_lightcurve}), and the double peak profile in its BAT light curve implies the photospheric radius expansion at its initial stage. Hence, 
in order to give a more accurate estimate to its fluence and effective duration, we fitted the fluxes of B2010 with a fixed peak flux equal to the Eddington flux $F_{\rm Edd}$.  All best-fitting parameters and derived properties are shown in Table~\ref{table_result}.

\begin{figure}[!htbp]
\plotone{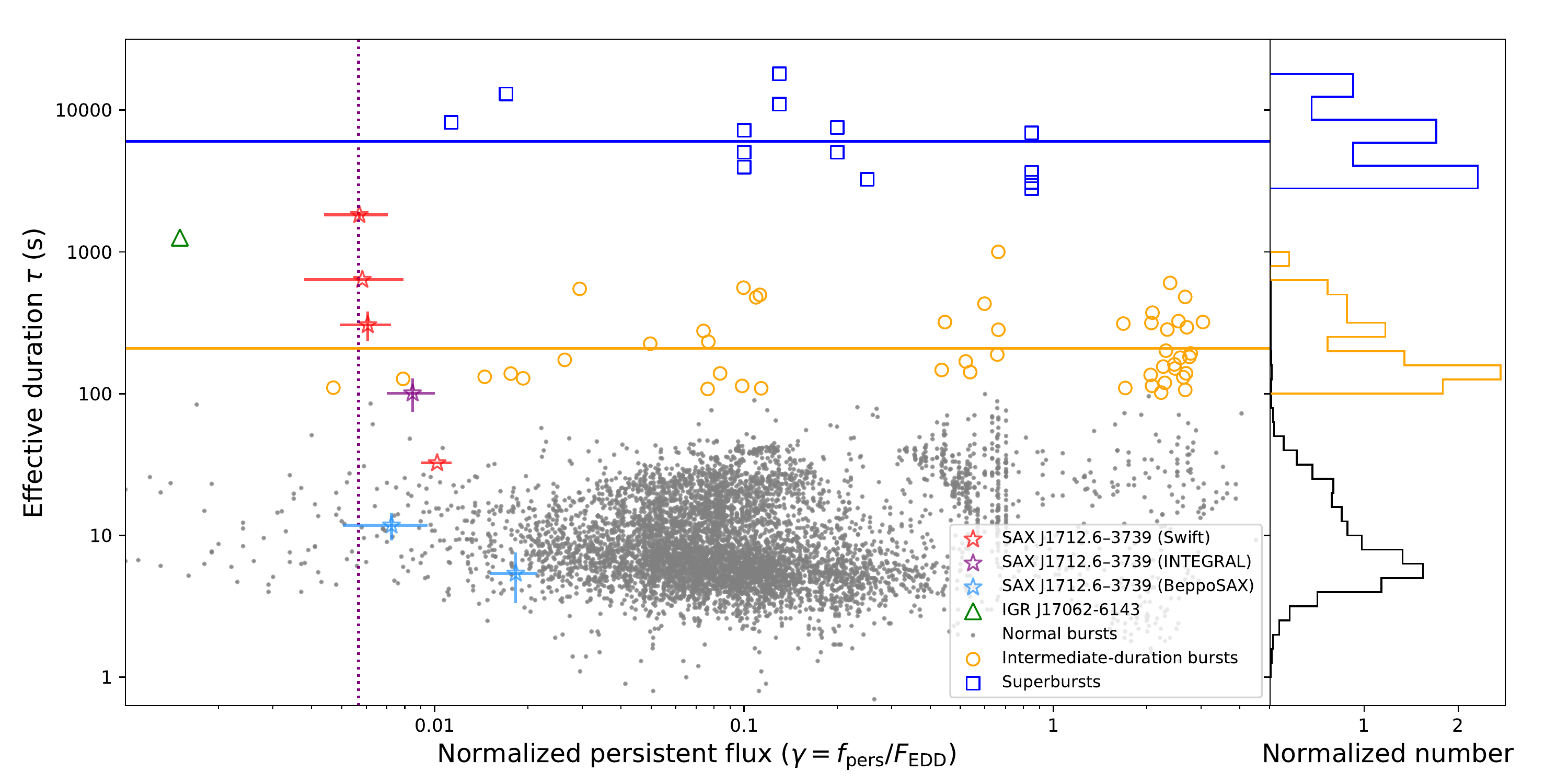}
\caption{
The effective duration against the normalized persistent flux for thermonuclear bursts. The measurements of the normal bursts and intermediate-duration bursts ($\tau\geq 100$~s here) are contributed by Multi-INstrument Burst ARchive (MINBAR, \citealt{MINBAR+2020}), and the measurements of superbursts are obtained from \cite{zand+2017,Zand+etal+2004,Kuulkers+2004,Keek+etal+2008,Kuulkers+etal+2010,Serino+etal+2012}. Additionally, the 2015 long burst observed from IGR J17062--6143 is also shown in the figure \citep{Keek+etal+2017}. The bursts of SAX J1712.6--3739 observed by BeppoSAX, INTEGRAL and \emph{Swift} are marked with cyan, purple and red, respectively. 
The vertical purple dotted line indicates the long-term average persistent flux derived from \emph{Swift}/BAT monitoring data. The orange and blue solid lines indicate the average effective duration for intermediate-duration bursts and superbursts, respectively. Furthermore, we also showed the normalized distribution functions in the right panel for the three main burst branches, respectively.
\label{figure_TauVsGamma}}
\end{figure}

%From the best-fitting models, the flux corresponding to Eddington luminosity is $F_{\rm Edd}= 1.07 ^{+0.21}_{-0.18} \times 10^{-7}\rm \,ergs\,s^{-1}\,cm^{-2}$. 
%We showed the best-fitting parameters, derived fluences and effective durations in the Table~\ref{table_result}.

We plotted the effective duration $\tau$ against the normalized persistent flux ($\gamma=f_{\rm pers}/F_{\rm Edd}$) for the bursts from SAX J1712.6--3739 in Figure~\ref{figure_TauVsGamma},. For comparison, the measurements of bursts in the Multi-INstrument Burst ARchive (MINBAR, \citealt{MINBAR+2020}) and of superburst in literature \citep{zand+2017,Zand+etal+2004,Kuulkers+2004,Keek+etal+2008,Kuulkers+etal+2010,Serino+etal+2012} are also shown in the figure. Similar to \cite{Falanga+etal+2008}, we labeled the MINBAR bursts with $\tau \geq 100$~s as ``intermediate-duration burst''. It is worth noting that, the measurements of two bursts of SAX J1712.6--3739 observed by BeppoSAX/WFC in 1999 \citep{Cocchi+etal+2001} and 2000, were published by MINBAR. We highlighted these two bursts in the figure. Additionally, we also showed the measurements for the intermediate-duration burst of SAX J1712.6--3739 observed by INTEGRAL/JEM-X on Feb. 20, 2018 \citep{Alizai+etal+2020}.
 Notice that, their normalized persistent fluxes here were calculated based on 10-day average (RXTE/ASM or \emph{Swift}/BAT) rates prior to the burst and then normalized with Eddington flux $F_{\rm Edd}$ we derived above \citep{Krimm+etal+2013}.
 
The effective duration of B2010 is consistent with the normal X-ray bursts (see Figure~\ref{figure_TauVsGamma}). B2011 corresponds to the distribution of intermediate-duration bursts, while B2014 locates in the gap between superbursts and intermediate-duration bursts.
B2014 is similar but stronger than the day-long burst observed from IGR J17062--6143 in 2015 ($\tau\approx 1300$~s, \citealt{Keek+etal+2017}). Hence, the 2014 burst event of SAX J1712.6--3739 could be known strongest intermediate-duration burst or weakest superburst.

%In the view of effective durations, the 2010 burst corresponds to normal X-ray burst,  the 2011 burst is comparable to intermediate-duration bursts, while the 2014 burst is further stronger than the common intermediate-duration bursts but weaker than superbursts.

\subsection{The 2018 May burst}

\begin{figure}
\plotone{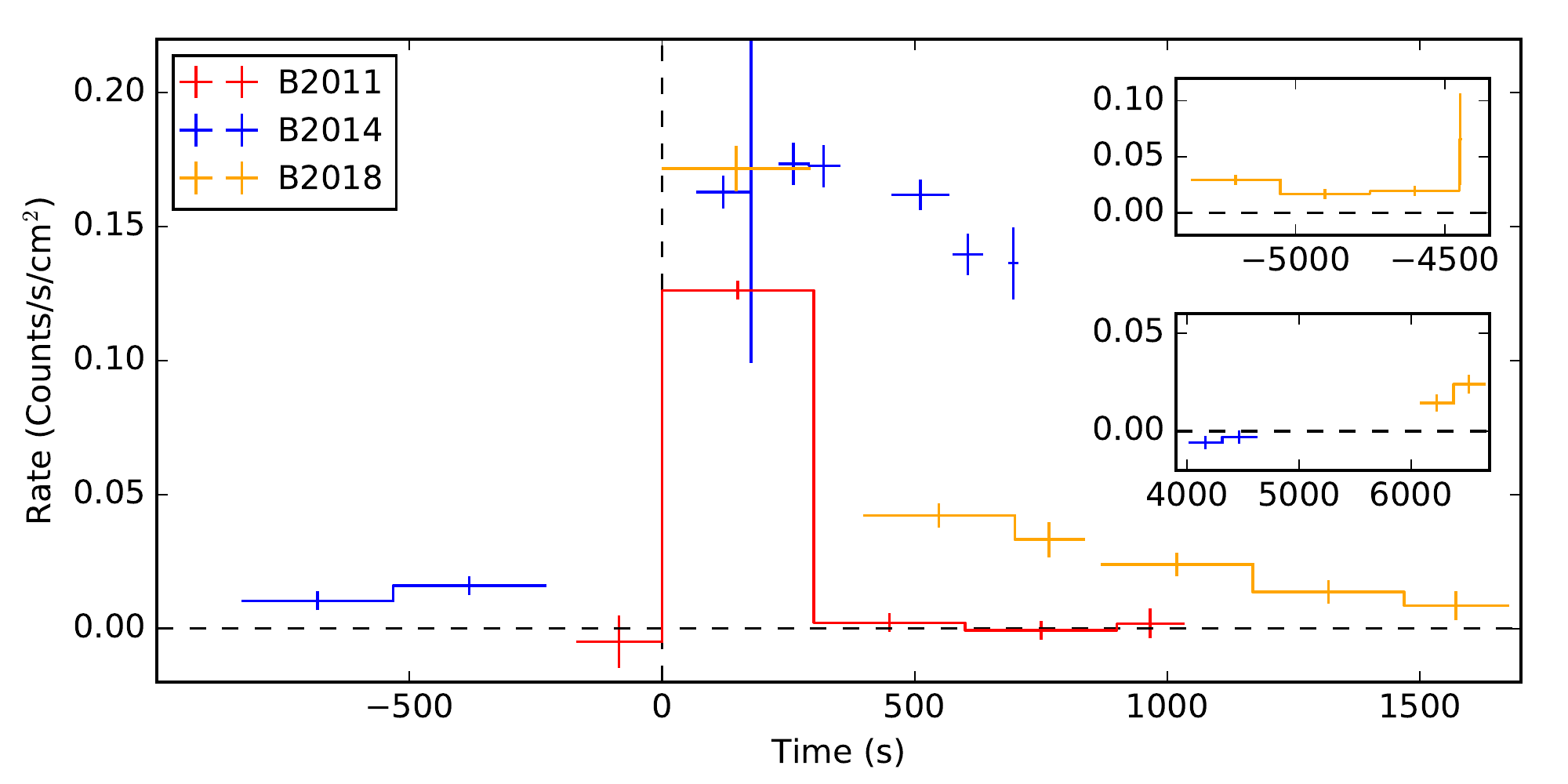}
\caption{The BAT light curves of the 2011, 2014, and 2018 burst events in the energy band of 15--100 keV with typical time bins of 300s. Among which, the light curve of the 2011 burst event was produced from the BAT event data, and the light curves of B2014 and B2018 were produced from BAT survey data. 
The start point of the 2018 burst event was set to UTC 2018 May 08 09:17:48, and the start point of the 2014 burst event was set to the time when its main burst started.
Because the observations of the 2014 and 2018 burst events were both interrupted by the slew of \emph{Swift}, we also show the light curves before and after the bursts in the right of the figure.  
\label{figure_compare}}
\end{figure}

As introduced in the Section~\ref{sec:data}, the 2018 burst event triggered neither BAT event-data recording nor XRT observation \citep{Lin+Yu+2018+atel}, thus we investigated this event by only BAT survey data.
We plotted the BAT light curve of B2018 in the Figure~\ref{figure_compare}, and also plotted the BAT light curves of B2011 and B2014 in similar time bins.
As Figure~\ref{figure_compare} shows, B2018 was already detected at the apparent peak count rates, $\approx 0.17 \rm \,counts\,s^{-1}\,cm^{-2}$, before SAX J1712.6--3739 entered the FoV of the \emph{Swift}/BAT. 
Hence, the duration of B2018 should be significantly longer than the duration of the \emph{Swift}/BAT observation ($ \approx$1700~s). The start point $T_0$ of the 2018 burst event was set to UTC 2018 May 08 09:17:48 when \emph{Swift} just slewed to the direction of ESO 324–24 where SAX J1712.6–3739 can be seen from off-axis direction.
We have also checked the BAT rate files and attitude files corresponding to B2018.
Until the first survey observation detected B2018, \emph{Swift} was slewing, implying that we cannot reproduce the source light curve by the rate data like what we did for B2014.
Based on the last survey observation (ended at 08:03:40 UTC) which covers the position of SAX J1712.6--3739, the duration of missing part of B2018 should not be longer than 4448 seconds. 
We also investigated the MAXI observations for the region of SAX J1712.6--3739, but no MAXI observation covered SAX J1712.6--3739 during the observational gap (08:03:40 to 09:17:48 UTC).

Due to the limit of observational time coverage and energy band, the natures of B2018 cannot be precisely measured.
It is worth noting that, B2014 has shown a flat profile in its initial phase, which may correspond to the PRE process. From available \emph{Swift}/BAT observations, B2018 showed a peak count rate consistent with the ``flat'' stage of B2014. If B2018 experienced a similar ``flat'' in its initial phase, the fluence for the missing part is not higher than $ F_{\rm Edd}\times 4448~{\rm s}  \approx 4.8\times 10^{-4}{\rm~ergs\,cm^{-2}}$.
By additionally assuming that the burst bolometric flux is proportional to the observed BAT count rate (15--100 keV), we roughly estimated the lower limit for its fluence and effective duration.
As shown in Table~\ref{table_result} and Figure~\ref{figure_TauVsGamma}, B2018 is much stronger than B2011 and seems to be comparable to B2014.

\section{Discussion} \label{sec:discussion}
\subsection{The natures of SAX J1712.6--3739}
\label{sec:basic_natures}
By assuming that B2011 and B2014 both reached the Eddington luminosity, the distance of SAX J1712.6--3739 can be estimated by the bolometric peak flux \citep{Kuulkers+etal+2003}.
The Eddington luminosity, measured by an observer at infinity, can be estimated as 
\begin{equation}
L_{\rm Edd, \infty}=\frac{4\pi {c}{\rm G}M_{\rm NS}}{\kappa(1+z)}=2.7\times10^{38}\,(\frac{M_{\rm NS}}{1.4 \rm \,M_\odot})\,(1+X)^{-1}\,(\frac{1+z}{1.31})^{-1}\rm\,ergs\,s^{-1}
\end{equation}
 \citep{Lewin+etal+1993,Galloway+etal+2008}, where $M_{\rm NS}$ is the mass of neutron star, $X$ represents the hydrogen fraction and $1+z=(1-2GM_{\rm NS}/R_{\rm NS}c^2)^{-1/2}$ is the gravitational redshift at the surface of neutron star.
For most common neutron star mass and radius, $R_{\rm NS}=10$~km and $M_{\rm NS}=1.4$~$\rm M_\odot$, the gravitational redshift is given as $1+z=1.31$. The mass measurement of neutron star in radio pulsar binaries suggested that neutron star masses are consistent with a narrow Gaussian distribution $M_{\rm Ns}=1.35\pm0.04$~$\rm M_\odot$ \citep{Thorsett+Chakrabarty+1999}. For a neutron star with a mass of 1.4~$\rm M_\odot$, its radius is most likely in the range from 10 to 14~km \citep{Steiner+etal+2018}, thus the uncertainty of gravitational redshift is only $\approx$3\% 
 %(by assuming the radius following a uniform distribution function within the range)
 . Even if the neutron star in X-ray binary spans a very large mass range, such as 1--2~$\rm M_\odot$ (see \citealt{Alsing+etal+2018}), the uncertainty of gravitational redshift still does not exceed 10\%.
 Therefore, the gravitational redshift term $1+z$ just introduces a small systematic error to the estimate of $L_{\rm Edd, \infty}$ (see also \citealt{Galloway+etal+2008}).
 
The distance to bursters can be estimated as $d=(L_{\rm Edd}/4\pi \xi_{\rm b} F_{\rm Edd})^{1/2}$, where $\xi_{\rm b}$ is the anisotropy factor of burst radiation due to the block or reflection of the disk (see \citealt{He+Keek+2016}). The anisotropy factor dominantly depends on the inclination and shape of the accretion disk. For a standard flat disk, $\xi_{\rm b}^{-1}$ decreases from $\approx$1.5 to 0.5 with the increase of the inclination angle (e.g. $\xi_{\rm b}^{-1}=\frac{1}{2}+\cos\theta$  for a flat Lambertian surface). For other shapes of accretion disk (e.g. the triangular disk), $\xi_{\rm b}^{-1}$ increases to $\approx$2 if its inclination angle is low enough or decreases to 0 as its inclination angle is very high \citep{He+Keek+2016}. However, neither the inclination nor the shape of disk can be well constrained for SAX J1712.6--3739, which leads high uncertainty into the estimate of its distance. \cite{zand+etal+2019} suggested that the inclination angle of SAX J1712.6--3739 is less than ${60}^{\circ}$ since the dips or eclipses from SAX J1712.6--3739 have not been seen yet. Therefore, $\xi_{\rm b}^{-1}$ should be in the range of 1 to $\approx$1.5 by assuming a flat disk; for the other disk shapes, $\xi_{\rm b}^{-1}$ varies but is still in the range of $\approx$1 
to $\approx2$ (see Figure~11 of \citealt{He+Keek+2016}). It is worth noting that, the large-amplitude light curve variabilities have been observed in the 2011 burst and the 2014 burst of SAX~J1712.6--3739, which can be explained by 
rapidly changing anisotropy factor \citep{He+Keek+2016}. However these dramatic variabilities will not affect the estimate of distance or burst fluence, because they appeared after PRE-phase (see Figure~\ref{figure_flux}) and did not change the cooling trend of burst (see also \citealt{Degenaar+etal+2018,zand+etal+2019} ) .

By assuming the the neutron star mass $M_{\rm NS}=1.4~{\rm M_\odot}$ and the gravitational redshift $1+z=1.31$ (with an uncertainty of 3\%), the distance of SAX J1712.6--3739 is $4.60\pm0.12\,{\rm kpc}\times\xi_{\rm b}^{-\frac{1}{2}}(1+X)^{-\frac{1}{2}}$. Hence, for a flat Lambertian disk with inclination angle less than ${60}^{\circ}$, 
the distance of SAX J1712.6--3739 is 3.5--4.3~kpc for a solar composition atmosphere, or 4.6--5.6~kpc for a pure helium atmosphere. As an UCXB candidate \citep{Zand+etal+2007}, the atmosphere of SAX J1712.6--3739 is more likely to be hydrogen-poor. Furthermore, even if the accreted fuel is hydrogen-rich, the bursts are still possible to reach the Eddington limit of pure helium atmosphere, since the intensive radiation pressure can expel or dilute the hydrogen layer and then the atmosphere becomes dominated by helium (see \citealt{Bult+etal+2019,Galloway+etal+2006} ). Therefore, the distance of SAX J1712.6--3739 is most likely to be 4.6--5.6~kpc.
Since the inclination,  neutron star mass and radius are highly uncertain, the distance of SAX~J1712.6--3739 cannot be tightly constrained by the measurements of PRE bursts yet.

%
%Nevertheless, we are convinced that the distance of SAX~J1712.6--3739 is not further than $\approx$6.5~kpc.

According to the bolometric peak flux ($\approx5\times10^{-8} \rm \,ergs\,s^{-1}\,cm^{-2}$) extrapolated from burst peak intensity of the 1999 burst event observed by BeppoSAX/WFC (2--28 keV), \cite{Cocchi+etal+2001} suggested that the distance of SAX~J1712.6--3739 is $\approx7$~kpc (not correcting for anisotropy or gravitational redshift). However, MINBAR recently gave a much higher bolometric peak flux ($1.6\pm0.5\times10^{-7} \rm \,ergs\,s^{-1}\,cm^{-2}$) for this burst event based on the time-resolved spectroscopy of BeppoSAX/WFC \citep{MINBAR+2020}. The intermediate-duration burst, detected by INTEGRAL/JEM-X (3--35~keV) on Feb. 20, 2018, reached a bolometric peak flux of $1.1\pm0.2\times10^{-7} \rm \,ergs\,s^{-1}\,cm^{-2}$ \citep{Alizai+etal+2020}, consistent with our measurements of Eddington flux. These new measurements support that the previous distance of SAX~J1712.6--3739 is very likely to be overestimated.
%The updated distance is lower than the previous estimate by maximum flux of the 1999 burst event in the BeppoSAX observations \citep{Cocchi+etal+2001}, because \cite{Cocchi+etal+2001} did not take gravitational redshift into account and \emph{Swift}/XRT captured much higher flux in its recent bursts. 

The distance of SAX J1712.6--3739 should be similar to those closest UCXBs, e.g. 4U 0614+09 and 4U 1246--58 \citep{zand+etal+2008,Kuulkers+etal+2010,Lin+Yu+2018}, but the column density of SAX J1712.6--3739 is higher by an order of magnitude, which may be related to the bow shock H$\alpha$ nebula \citep{Wiersema+etal+2009}. The edge-brightened cone-shaped bow shock, observed around SAX J1712.6--3739, could be powered by two continuously replenished jets from a LMXB \citep{Heinz+etal+2008}.
Noted that, with updated distance, the diffuse region of the H$\alpha$ nebula would be much smaller, and the leading edge of this region corresponds to $\sim2 \times 10^{18}~{\rm cm}$ if $d=4.6~{\rm kpc}$. Hence, the energy budget of the jet should be also much lower (e.g. $\lesssim 10^{36}~{\rm ergs\,s^{-1}}$), and is similar to the observed jet power of UCXB 4U 0614+091 \citep{Migliari+etal+2006}.

On the other hand, the updated distance, the visual magnitude of $m=23.93$ measured by EFOSC2 \citep{Wiersema+etal+2009} and the predicted visual extinction $A_{\rm V}=7.3$ by \cite{Zand+etal+2007} imply that the absolute magnitude of SAX J1712.6--3739 is $M=3.3-5\log(d/4.6~{\rm kpc})$, which agrees the UCXB nature of SAX J1712.6--3739 \citep{Paradijs+McClintock+1994,Zand+etal+2007}.
 
 For searching the UCXB candidates from persistent LMXBs, \cite{Zand+etal+2007} had estimated the the persistent bolometric flux for 40 X-ray bursters. 
 In their works, the estimated bolometric flux of of SAX J1712.6--3739 was about $F_{\rm per}=2.6\times10^{-10}\,\rm ergs\,s^{-1}\,cm^{-2}$ from RXTE/PCA observations and $F_{\rm per}=6.7\times10^{-10}\,\rm ergs\,s^{-1}\,cm^{-2}$ from long-term RXTE/ASM monitoring observations. 
 They also noted that its flux can reached $1.6\times 10^{-9}\,\rm ergs\,s^{-1}\,cm^{-2}$ at its active state around 2007. 
 The \emph{Swift}/BAT monitoring data shows that the weighted average count rate of SAX J1712.6--3739 was $1.41\pm0.02\times 10^{-3}~\rm counts\,s^{-1}\,cm^{-2}$ ($\approx 6.4~ \rm mCrab$), which is consistent with the monitoring results from INTEGRAL \citep{Fiocchi+etal+2008}.
The count rate corresponds to a 0.1--200 keV flux of $F_{\rm per}=6.1\times10^{-10}\,\rm ergs\,s^{-1}\,cm^{-2}$ by assuming a Crab spectrum \citep{Kirsch+etal+2005+CRAB}.
Thus the average accretion rate of SAX J1712.6-3739 is $ \dot{M}=\xi_{\rm p}F_{\rm per}/\xi_{\rm b}F_{\rm Edd}\times \dot{\rm M}_{\rm Edd}\approx 0.6\%\,\rm \dot{M}_{ Edd} \times (\xi_{\rm p}/\xi_{\rm b})$, where $\xi_{\rm p}$ is the anisotropy factor for persistent flux. By assuming the inclination angle less than $60^{\circ}$, the average accretion rate can be corrected to 0.4--$0.7\%\,\rm \dot{M}_{ Edd}$ (see Figure~12 of \citealt{He+Keek+2016}),
which is relatively low in the bursters \citep{Falanga+etal+2008}. The accretion rate supports its persistent UCXB nature under the disk instability model \citep{Zand+etal+2007}.

\subsection{The fuel of the long thermonuclear bursts}
\label{sec:disscussion:fuel}
As shown in Table~\ref{table_result}, the integrated bolometric burst energy, measured by an observer at infinity, can be estimated from the burst fluence as $E_{\rm b,\infty}=4\pi d^{2}\xi_{\rm b}f_{\rm b}$. Then ignition column depth can be estimated from the burst energy, as
\begin{equation}
y=\frac{E_{\rm b,\infty}(1+z)}{4\pi R_{\rm NS}^2 Q_{\rm nuc}}=2.09\times10^{10} {\rm\,g\,cm^{-2} }(\frac{f_{\rm b}}{10^{-4}~{\rm ergs\,cm^{-2}}})(\frac{d}{4.6\,{\rm kpc}\times\xi_{\rm b}^{-\frac{1}{2}}})^{2}(\frac{Q_{\rm nuc}}{1.31~\rm MeV\,nucleon^{-1}})^{-1}(\frac{R_{\rm NS}}{10~{\rm km}})^{-2}(\frac{1+z}{1.31}),
\end{equation}
where $R_{\rm NS}=10 {\rm km}$ is the radius of neutron star, and $Q_{\rm nuc}$ represents the nuclear energy release from the fuel \citep{Galloway+etal+2008}.
%For carbon burning into iron-peak elements in the superbursts, the energy release is $Q_{\rm C}\approx1\,{\rm MeV\,nucleon^{-1}}\approx \rm 10^{18}\,ergs\,g^{-1}$. Otherwise,  
The energy release appreciate for H/He burning to heavier elements is $Q_{\rm H/He}=1.31+6.95X-1.92X^2\,{\rm MeV\,nucleon^{-1}}$ \citep{Goodwin+etal+2019}. Thus for a pure helium burst ($X=0$), which is usually used to account for the intermediate-duration burst, the energy release approaches $Q_{\rm He}=1.31\,{\rm MeV\,nucleon^{-1}}$.

The UCXB nature implies that its atmosphere is hydrogen-poor, thus the thermonuclear bursts from SAX J1712.6--3739 are very likely to be pure helium bursts.
We showed the ignition column depths for bursts in Table~\ref{table_result}.
Since B2014 and B2018 both have a higher radiative energy and a longer effective duration than common intermediate-duration bursts, thus we discuss below whether they have the same origin as the superbursts.

\cite{Cumming+etal+2006} found that the carbon mass fraction of the superburst fuel is 15--30\% by comparing the observed superburst light curves to their cooling models. The carbon in the superbursts is mainly produced by 3$\alpha$ process rather than accreted from donors. However, the high temperature during unstable nuclear burning allows the reactions to run up to iron-peak elements \citep{Strohmayer+Brown+2002,Galloway+Keek+2017}. In the simulations, the net carbon production by H/He bursts is not more than a mass fraction of 5\% \citep{Woosley+etal+2004}. Hence, only stable H/He burning can produce enough carbon fuel supplying for superbursts. In theory, the known stable H/He burning regimes are $\dot{M}\approx 10\%\,\dot{\rm M}_{\rm Edd}$ \citep{Keek+Heger+2016}  and $\dot{M} \gtrsim \dot{\rm M}_{\rm Edd}$ \citep{Heger+etal+2007}. However, the average accretion rate of SAX J1712.6--3739, $ 0.4-0.7\%\,\rm \dot{M}_{ Edd}$, is far lower than these burning regimes, implying that SAX J1712.6--3739 lacks of efficient mechanism to accumulate substantial fraction of carbon in current theories.

On the other hand,  in observations, the superbursts now known were all produced above the accretion rate of $10\%\,\rm \dot{M}_{ Edd}$, except those superbursts in 4U 0614+091 \citep{zand+2017}. The 2005 superburst of 4U 0614+091 lasted for 2.4 hours at least while the accretion rate of 4U 0614+091 was only $1\%\,\rm \dot{M}_{ Edd}$\citep{Kuulkers+etal+2010,Lin+Yu+2018}, which is difficult to be explained by carbon burning. \cite{Kuulkers+etal+2010} suggested that the observed superbursts can also be caused by ignition of a thick He layer. This explanation might also apply to the energetic bursts in SAX J1712.6--3739.

In theory, stronger pure helium burst is produced at lower accretion rate (see \citealt{Cumming+etal+2006}).
For the very low accretion rate of SAX J1712.6--3739, the ignition column depth for pure helium bursts will reach several $10^{10} \rm \,g\,cm^{-2}$, which is comparable to the estimated ignition column depth of B2014 and B2018 (Table~\ref{table_result} and Figure~\ref{fig:ignition_rate}). Therefore, B2014 and B2018 may both be deep helium bursts \citep{Galloway+Keek+2017,zand+etal+2019}.

\subsection{The diverse durations of thermonuclear bursts}

A few thermonuclear bursts of SAX J1712.6--3739 had been discovered by X-ray detectors so far. Interestingly, the duration of these bursts varied over a very wide range, in which the integrated radiative energies were from a few $10^{39} \rm ergs$ to several $10^{41} \rm ergs$. 
It is worth noting that, 4U 0614+091 has also shown diverse bursts in which the radiative energies span by two orders of magnitude.
\cite{Kuulkers+etal+2010} inferred that the different ignition depth of helium layer caused by the variable accretion rates could be responsible for the diverse durations.

\begin{figure}[!htbp]
\centering
    \includegraphics[width=0.6\textwidth]{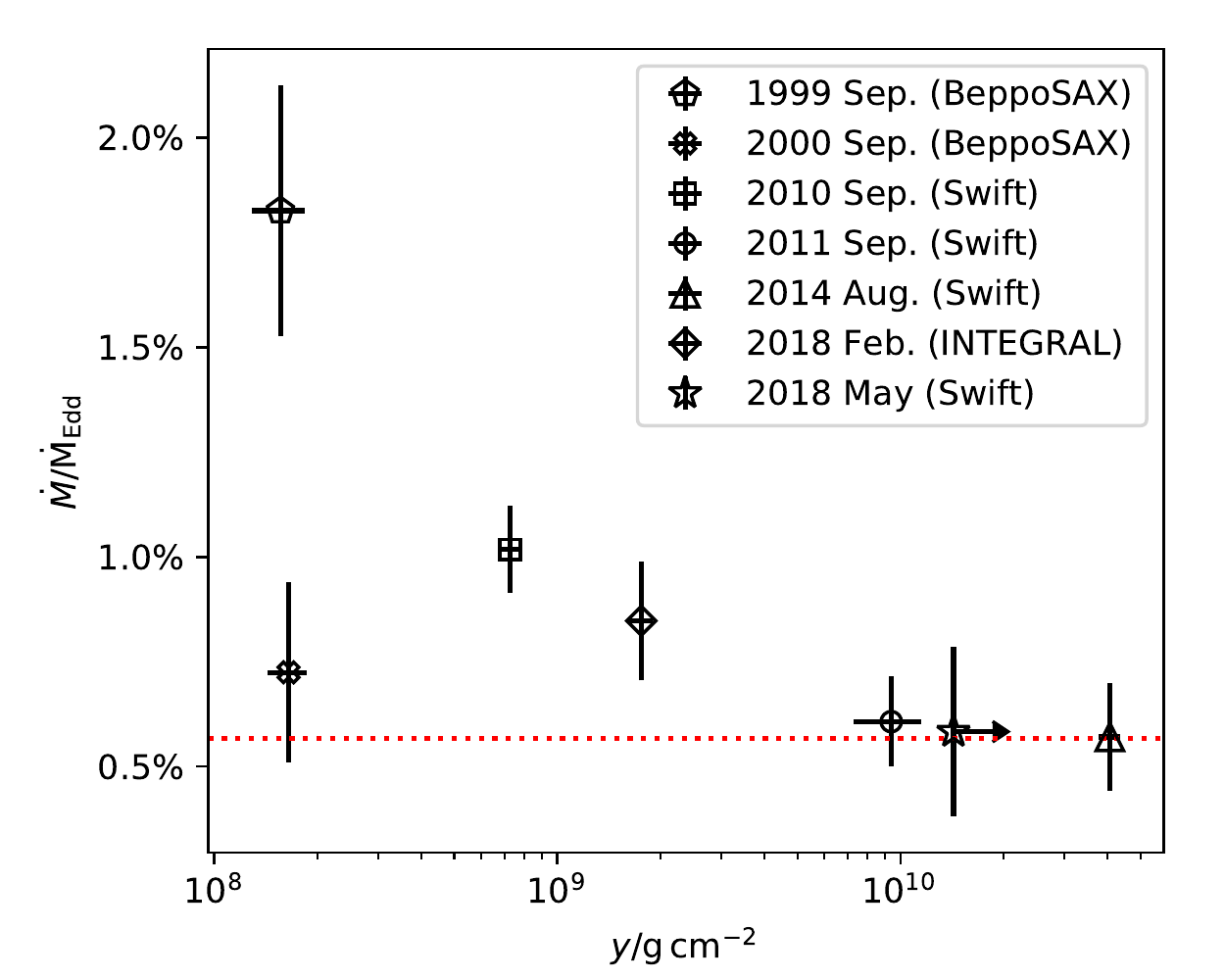}
    \caption{
     The relation between the mass accretion rates and the ignition column depths for the thermonuclear bursts in SAX J1712.6--3739. The mass accretion rates are the 10-day average rates prior to the burst, by assuming that the persistent spectra are similar to the Crab spectrum and the Eddington flux is set to $F_{\rm Edd}=1.07 \times 10^{-7} \rm\,ergs\,s^{-1}\,cm^{-2}$. The ignition column depths are calculated by assuming pure helium fuel ($Q=1.31\,{\rm MeV\,nucleon^{-1}}$), distance of $4.60\pm0.12\,{\rm kpc}\times\xi_{\rm b}^{-\frac{1}{2}}$ and neutron star radius $R_{\rm NS}=10$~km.
     Notice that, the error bars of accretion rates here represent the uncertainty contributed by \emph{Swift}/BAT (or RXTE/ASM)  count rates,
     and the error bars of ignition column depths were contributed by the uncertainty of fluences and peak fluxes. 
     %We showed the ignition depths for the 2010 burst event (empty square) in two sets of different best-fitting parameters.
     The arrow indicates that the ignition column depth for the 2018 burst event here is a lower limit.
     The red dotted line represents the long-term average persistent mass accretion rate derived from \emph{Swift}/BAT monitoring data.}
    \label{fig:ignition_rate}
\end{figure}

The Figure~\ref{fig:ignition_rate} shows the relation between the ignition column depth and the mass accretion rate prior to burst.
Noted that, the conversion from \emph{Swift}/BAT (or RXTE/ASM) count rates to mass accretion rates introduces more uncertainty, depending on the shape of energy spectrum and radiative efficiency,
while the error bars of accretion rates in our figure just represent the uncertainty from count rates.
Similarly, the estimate of ignition column depth is based on the assumption that the radius of neutron star $R_{\rm NS}$ and the nuclear energy release $Q_{\rm nuc}$ are both set to constant instead of introducing additional uncertainty. 

These points roughly agree with the ignition models in Figure~18 of \cite{Cumming+etal+2006}, except the 2000 burst event observed by BeppoSAX/WFC \citep{MINBAR+2020}. 
These bursts follow a broad trend that the ignition column depth is higher for a lower mass accretion rate, but the bursts observed by BeppoSAX show similar ignition depths at distinct accretion rates. Therefore, it is still difficult to conclude that the ignition column depth depends on the mass accretion rate prior to the burst. It is worth noting that, B2011, B2014 and B2018 have very similar mass accretion rates, but their ignition column depths span by an order of magnitude at least. This may indicate that, at lower accretion rates, the ignition column depths are more sensitive to the change of accretion rates (see \citealt{Cumming+etal+2006}). Further test requires more sensitive observations for the pre-burst fluxes and better observation coverage for the bursts.

\subsection{The ``precursor'' in the 2014 burst event}

As introduced in Section~\ref{sec:intro}, the ``precursor'' of thermonuclear bursts usually implies extreme PRE process in the early phase of burst. 
Hence, the ``precursor'' in B2014 could also be caused by the decrease of the effective temperature, which shifted the X-ray emission out of the observational energy band of \emph{Swift}/BAT. However, because the BAT data quality  was poor and the XRT observation did not cover the initial stage of B2014, further identification for the ``precursor''  cannot be performed. Therefore, we also discuss another hypothesis below. 

In the Section~\ref{sec:light_curve}, we have shown that the BAT count rates, duration and energy spectrum of ``precursor'' are similar to those of the 2010 burst. Due to the BAT energy band of 15--200 keV, the weak burst events in BAT light curves are actually very energetic in the observations at soft X-ray energy band. 
Hence, in the soft X-ray band, the main radiative band for X-ray bursts, the duration of ``precursor'' should be far longer than $ \approx15$~s and the ``precursor'' could be a ``short'' burst like the 2010 burst, although the detailed mechanism for such burst event is not clear yet. B2014 may be similar to the superburst event of 4U 1820--30 in 1999 September (hereafter S1999, \citealt{Strohmayer+Brown+2002}), in which a helium flash occurred just prior to the onset of the superburst.

It is worth noting that, S1999 is the only superburst to show large-amplitude variabilities in its X-ray light curves, while other large-amplitude variabilities in burst light curves were all discovered in the intermediate-duration bursts \citep{Degenaar+etal+2018,zand+etal+2019}.
 Although the accretion rate of 4U 1820--30 is much higher than that of all other UCXBs \citep{Zand+etal+2007,Lin+Yu+2018}, the superburst in 4U 1820--30 is still difficult to be explained by carbon burning \citep{Strohmayer+Brown+2002}. Because the 11-minute orbital period of 4U 1820--30 requires that its donor is a low-mass helium dwarf, the helium-rich accretion environment does not allow any stable burning regimes below Eddington accretion rate (see \citealt{Galloway+Keek+2017}). In this view, S1999 may be a very long intermediate-duration burst rather than a superburst, like the observed longest bursts in SAX J1712.6--3739 and 4U 0614+091. 

\section{Summary} \label{sec:summary}
By measuring the fluence and the effective duration for four energetic bursts in SAX J1712.6--3739 observed with \emph{Swift}, we determined that the 2010 burst event corresponds to normal X-ray bursts,
the 2011 burst event is consistent with intermediate-duration bursts while the 2014 and 2018 burst events are more energetic than conventional intermediate-duration bursts but less energetic than the superbursts. 
If the current model that predicts no carbon production in the bursters under very low accretion rates is correct, the 2014 and the 2018 bursts should be deep helium bursts.

Since the higher peak fluxes (typically $F_{\rm peak} \gtrsim \rm 10^{-7}\,ergs\,s^{-1}\,cm^{-2}$) in thermonuclear bursts have been measured and the gravitational redshift effect is also taken into account, the distance of SAX J1712.6--3739 is essentially much closer than previous estimate of 7~kpc. Hence, the inferred absolute magnitude and average accretion rate agree its persistent UCXB nature. However, its inclination, neutron star mass and radius are still highly uncertain, the anisotropy factor and thus the distance of SAX J1712.6–3739 cannot be tightly constrained yet.

The thermonuclear bursts of SAX J1712.6--3739 have shown a very wide range of duration.
These bursts are possibly ignited at different depth of helium layer caused by the variable accretion rates, but the poor observational data provides only weak support to this hypothesis.
The burst ignition column depths and accretion rates prior to the burst roughly agree with the ignition models for pure helium bursts \citep{Cumming+etal+2006}, except the 2000 burst event observed by BeppoSAX/WFC. Although the performances of these bursts are very different, their physical origins are likely to be consistent.

Furthermore, the observational evidence is not sufficient to confirm whether the ``precursor'' of the 2014 burst was caused by PRE process or not. In current BAT observations, the ``precursor'' is very similar to the 2010 burst in their X-ray light curves and energy spectra. If the ``precursor'' is another ``short'' burst, the 2014 burst will meet a similar scenario to the superburst event of 4U 1820--30 in 1999 September. The detailed mechanism for a short burst prior to a very long burst is still unclear.

\acknowledgments
\vspace{5mm}
We are grateful to Xiaofeng Wang (THCA), Jean J. M. in't Zand (SRON) and Shangyu Sun (SHAO) for supports and very useful discussions.
We thank \emph{Swift} Guest Observer Facilities at NASA Goddard Space Flight Center for providing \emph{Swift}/XRT products and \emph{Swift}/BAT transient monitoring results.
This research has made use of MAXI data provided by RIKEN, JAXA, and the MAXI team, and RXTE/ASM data provided by RXTE/ASM team.
W.Y. would like to acknowledge the support by the National Program on Key Research and Development Project (Grant no. 2016YFA0400804), by the National Natural Science Foundation of China under grant No. U1838203 and 11333005 and by the FAST fellowship, which is supported by Special Funding for Advanced Users, budgeted and administrated by Center for Astronomical Mega-Science, Chinese Academy of Sciences (CAMS). 

%\facilities{Swift(BAT and XRT)}
%\software{astropy } %\citep{2013A&A...558A..33A}

\iffalse

\fi

%\bibliographystyle{aasjournal}
%\bibliography{references}

\end{document}